\documentclass[12pt]{article} 
\usepackage{graphicx}
\RequirePackage{mathrsfs,multirow,morefloats,booktabs,subfigure,pdfpages,color,array,hhline,makecell,epstopdf,algorithm}
\RequirePackage[colorlinks,citecolor=blue,urlcolor=blue]{hyperref}
\RequirePackage{algpseudocode}
\algrenewcommand\algorithmicrequire{\textbf{Input:}}
\algrenewcommand\algorithmicensure{\textbf{Output:}}
\usepackage{amsmath,amssymb}
\RequirePackage{natbib}

\begin{document}

\title{\bf Mixed-SCORE+ for mixed membership community detection}
\author{Huan Qing\\
	Department of Mathematics, China University of Mining and Technology\\
	and \\
	Jingli Wang \\
	School of Statistics and Data Science, Nankai University}
\maketitle

\begin{abstract}
Mixed-SCORE is a recent approach for mixed membership community detection proposed by  \cite{mixedSCORE} which is an extension of SCORE \citep{SCORE}. In the note \cite{SCORE+},  the authors propose SCORE+ as an improvement of SCORE to handle with weak signal networks. In this paper, we propose a method called Mixed-SCORE+ designed based on the Mixed-SCORE and SCORE+, therefore Mixed-SCORE+ inherits nice properties of both Mixed-SCORE and SCORE+. In the proposed method, we consider $K+1$ eigenvectors when there are $K$ communities to detect weak signal networks. And we also construct vertices hunting and membership reconstruction  steps to  solve the problem of mixed membership community detection. 
Compared with several benchmark methods, numerical results show that Mixed-SCORE+ provides a significant improvement on the Polblogs network and two weak signal networks Simmons and Caltech, with error rates 54/1222, 125/1137 and 94/590, respectively. Furthermore, Mixed-SCORE+ enjoys excellent performances on the SNAP ego-networks.
\end{abstract}

	\noindent%
{\it Keywords:}  Mixed membership community detection; spectral clustering; Mixed-SCORE; SCORE+; weak signal network
\vfill

\section{Introduction}
Mixed membership community detection is a problem that has received substantial attentions, see \cite{MMSB, goldenberg2010a, mixedSCORE, GeoNMF, SPACL, MixedRSC, OCCAM}. In a mixed membership network, nodes may share among two or more communities. If a node only belongs to one community, we say this node is pure. For non-mixed membership community detection problem, all nodes are pure. In this paper, we focus on the study of mixed membership community detection.

Consider an undirected, un-weighted, no-loops network $\mathcal{N}$ and assume that there are $K$ disjoint blocks $V^{(1)}, V^{(2)}, \ldots, V^{(K)}$ where $K$ is assumed to be known in this paper. Let $A$ be its adjacency matrix such that $A_{ij}=1$ if there is an edge between node $i$ and $j$, $A_{ij}=0$ otherwise. 

This paper considers the degree-corrected mixed membership (DCMM) model \citep{mixedSCORE} which assumes that for each node $i$, there is a Probability Mass Function (PMF) $\pi_{i}=(\pi_{i}(1), \pi_{i}(2), \ldots, \pi_{i}(K))$ such that
\begin{align*}
	\mathrm{Pr}(i\in V^{(k)})=\pi_{i}(k), \qquad 1\leq k\leq K, 1\leq i\leq n.
\end{align*}
In this sense, DCMM model allows one node belongs to some certain communities with different probabilities.
By \cite{mixedSCORE}, under DCMM, we have $$\Omega=E[A]=\Theta \Pi P \Pi' \Theta,$$ where $\Theta$ is an $n\times n$ matrix whose $i$-th diagonal entry is the degree heterogeneity of node $i$, $P$ is a $K\times K$ matrix such that $Pr(A(i,j)=1|g_{i}=k,g_{j}=l)=\Theta(i,i)\Theta(j,j)P(g_{i},g_{j})$ (where $g_{i}$ denotes the community that $i$ belongs to). Therefore, given $(n, P, \Theta, \Pi)$, we can generate \footnote{For more details about how to generate $A$ under DCMM, please refer to \cite{mixedSCORE} and \cite{SCORE}.} a random adjacency matrix $A$ under the DCMM model. Let $\theta$ be the $n\times 1$ vector such that $\theta(i)=\Theta(i,i)$.
Let $\Pi$ be an $n\times K$ matrix such that its $i$-th row is $\pi_{i}$ for $1\leq i\leq n$. For the problem of mixed membership community detection, the chief aim is to estimate $\Pi$ with given $(A, K)$.

The Mixed-SCORE method \citep{mixedSCORE} is an extension of the SCORE method \citep{SCORE} to mixed membership community detection problem. As discussed in \cite{SCORE+}, traditional spectral clustering methods like SCORE, OCCAM \citep{OCCAM}, RSC \citep{RSC} can not deal with weak signal networks (defined in \cite{SCORE+}, and we redefined in our Algorithm) such as Simmons and Caltech \citep{traud2011comparing,traud2012social}. Therefore, \cite{SCORE+} proposed the SCORE+ as a simple improvement of SCORE to deal with weak signal networks. Some recent spectral clustering community detection methods proposed by  \cite{ NPCC, DRSC, ISC} can also successfully detect communities for weak signal networks. In this paper, we find that Mixed-SCORE also fails to detect Simmons and Caltech, which motivates us to design one approach which should successfully deal with mixed membership and weak signal networks.  Combining with Mixed-SCORE and SCORE+, we propose Mixed-SCORE+ as a refinement of Mixed-SCORE to weak signal networks, and it also can be deemed as an extension of SCORE+ to mixed membership networks. We list several important differences between Mixed-SCORE+ and Mixed-SCORE as well as SCORE+ as follows:
\begin{itemize}
	\item SCORE+ is for non-mixing community detection problem and it is designed based on the degree-corrected stochastic block model (DCSBM) \citep{DCSBM}, while Mixed-SCORE and Mixed-SCORE+ are for  the mixed membership community detection and designed based on the degree-corrected mixed membership (DCMM) model \citep{mixedSCORE}.
	\item Mixed-SCORE+ uses a regularized Laplacian matrix that is slightly different as the one used in SCORE+ and Mixed-SCORE.
	\item Mixed-SCORE+ has a threshold step while there is no such steps in SCORE+. However, when it turns to mixed membership community detection, there is also a threshold step  for mixed-SCORE.
	\item There are a vertices hunting (VH) step and a membership reconstruction (MR) step in Mixed-SCORE+ and Mixed-SCORE while there is no such steps in SCORE+.
	\item Mixed-SCORE+ applies the information of the leading $(K+1)$ eigenvectors and eigenvalues of a regularized Laplacian matrix for estimating $\Pi$ while Mixed-SCORE applies the leading $K$ eigenvectors of $A$. This enables that Mixed-SCORE+ can detect weak signal networks while Mixed-SCORE can not.
\end{itemize}

\section{The algorithm: Mixed-SCORE+}\label{sec2}
In this paper, for convenience, when we say ``leading eigenvalues'' or ``leading eigenvectors'', we are comparing the \emph{magnitudes} of the eigenvalues and their respective eigenvectors with unit-norm.

The details of Mixed-SCORE+ are presented in the following Algorithm.

\noindent\rule{14cm}{0.4pt}

\textbf{Mixed-SCORE+}. Input:  $A, K$, a ridge regularizer $\tau\geq 0$, two thresholds $t>0$ and $T_{n}>0$. Output: $\hat{\Pi}$.

\noindent\rule{14cm}{0.4pt}

$\bullet$ \texttt{SCORE+ step}:

1. Obtain the regularized graph Laplacian matrix by
\begin{align*}
	L_{\tau}=D_{\tau}^{-1/2}AD_{\tau}^{-1/2},
\end{align*}
where $D_{\tau}=D+\tau I$, $D$ is an $n\times n$ diagonal matrix whose  $i$-th diagonal entry is $D(i,i)=\sum_{j=1}^{n}A(i,j)$ (a good default $\tau$ is $\tau=0.1\frac{d_{\mathrm{max}}+d_{\mathrm{min}}}{2}$, where $d_{\mathrm{max}}=\mathrm{max}_{i}D(i,i), d_{\mathrm{min}}=\mathrm{min}_{i}D(i,i)$).

2. Asses the aforementioned ``signal weakness'' by $1-|\frac{\hat{\lambda}_{K+1}}{\hat{\lambda}_{K}}|$, and include an additional eigenvector for clustering if and only if
\begin{align*}
	1-|\frac{\hat{\lambda}_{K+1}}{\hat{\lambda}_{K}}|\leq t, \qquad (\mathrm{conventional~choise~of~}t~\mathrm{is~}0.1),
\end{align*}
where $\hat{\lambda}_{i}$ is the $i$-th leading eigenvalue of $L_{\tau}, 1\leq i\leq (K+1)$.

3. Let $M$ be the number of eigenvectors we decide in the last step (so either $M=K$ or $M=K+1$). Obtain the $n\times (M-1)$ matrix of entry-wise eigen-ratios by
\begin{align*}
	\hat{R}=[\frac{\hat{\eta}_{2}}{\hat{\eta}_{1}}, \frac{\hat{\eta}_{3}}{\hat{\eta}_{1}}, \ldots, \frac{\hat{\eta}_{M}}{\hat{\eta}_{1}}],\qquad \mathrm{where~}\hat{\eta}_{k}=\hat{\lambda}_{k}\hat{\xi}_{k}, 1\leq k\leq M,
\end{align*}
and $\hat{\xi}_{i}$ is the $i$-th leading eigenvector with unit-norm of $L_{\tau}, 1\leq i\leq (K+1)$.

4. Fixing a threshold $T_{n}$, define an $n\times (M-1)$ matrix $\hat{R}^{*}$ such that for all $1\leq i\leq n$ and $1\leq k\leq M$,
\begin{equation*}
	\hat{R}^{*}(i,k)=
	\begin{cases}
		\hat{R}(i,k), & \mathrm{if~} |\hat{R}(i,k)|\leq T_{n}, \\
		T_{n}, &\mathrm{if~} \hat{R}(i,k)>T_{n},\\
		-T_{n}, & \mathrm{if~} \hat{R}(i,k)<-T_{n},
	\end{cases}
\end{equation*}
where a good default $T_{n}$ is $\mathrm{log}(n)$.

$\bullet$ \texttt{Vertices Hunting (VH) step}:

5. Perform K-means clustering on the rows of $\hat{R}^{*}$ and obtain $K$ estimated cluster centers $\hat{v}_{1}, \hat{v}_{2}, \ldots, \hat{v}_{K} \in\mathcal{R}^{1\times (M-1)}$, i.e.,
\begin{align*}
	\{\hat{v}_{1}, \hat{v}_{2}, \ldots, \hat{v}_{K}\}=\mathrm{arg~}\underset{\hat{v}_{1},  \ldots, \hat{v}_{K}}{\mathrm{min}}\frac{1}{n}\sum_{i=1}^{n}\underset{\hat{v}\in\{\hat{v}_{1},\ldots, \hat{v}_{K}\}}{\mathrm{min}}\|\hat{R}^{*}_{i}-\hat{v}\|_{2}.
\end{align*}
Form the $K\times (M-1)$ matrix $\hat{V}$ such that the $i$-th row of $\hat{V}$ is $\hat{v}_{i}, 1\leq i\leq K$.

$\bullet$ \texttt{Membership Reconstruction (MR) step}:

6. Obtain the $K\times M$ matrix $\hat{V}_{*}$ by $\hat{V}_{*}=[\textbf{1}, \hat{V}]$, where $\textbf{1}$ is a $K\times 1$ vector with all entries being 1. Meanwhile, obtain an $n\times M$ matrix $\hat{R}^{*}_{*}$ by $\hat{R}^{*}_{*}=[\textbf{1}, \hat{R}^{*}]$, where $\textbf{1}$ is an $n\times 1$ vector with all entries being 1.

7. Project the rows of $\hat{R}^{*}_{*}$ onto the spans of $K$ rows of $\hat{V}_{*}$, i.e., compute the $n\times K$ matrix $\hat{Y}$ such that $\hat{Y}=\hat{R}^{*}_{*}\hat{V}_{*}'(\hat{V}_{*}\hat{V}_{*}')^{-1}$.

8. If there exists any node $i$ such that all entries of the $i$-th row of $\hat{Y}$ are negative, we set $\hat{Y}_{i}=-\hat{Y}_{i}$ (i.e., let all negative entries of $\hat{Y}_{i}$ be positive).

9. For $1\leq i\leq n, 1\leq k\leq K$, let $\hat{Y}(i,k)=\mathrm{max}(0, \hat{Y}(i,k))$.

10. Estimate $\pi_{i}$ by $\hat{\pi}_{i}=\hat{Y}_{i}/\|\hat{Y}_{i}\|_{1}, 1\leq i\leq n$.  Obtain the estimated membership matrix $\hat{\Pi}$ such that its $i$-th row is $\hat{\pi}_{i}, 1\leq i\leq n$.

\noindent\rule{14cm}{0.4pt}

Several remarks about Mixed-SCORE+ method are listed in order.
\begin{itemize}
	\item The regularized Laplacian matrix in Mixed-SCORE+ is slightly different from that of SCORE+, where we set $\tau=0.1 \frac{d_{\mathrm{max}}+d_{\mathrm{min}}}{2}$ instead of the $0.1d_{\mathrm{max}}$ in SCORE+ since such setting provides us with slightly better numerical results.
	\item In step 2, we measure the ``signal weakness'' slightly different as that in \cite{SCORE+}, where we use $1-|\frac{\hat{\lambda}_{K+1}}{\hat{\lambda}_{K}}|$ instead of the $1-\frac{\hat{\lambda}_{K+1}}{\hat{\lambda}_{K}}$ in \cite{SCORE+} since we find that the leading eigenvalues are measured by magnitude, which means that $\hat{\lambda}_{K+1}$ may have different sign as that of $\hat{\lambda}_{K}$.
	\item Similar as Mixed-SCORE,  in step 4, we need the threshold $T_{n}$ to guarantee the performances of Mixed-SCORE+. The default of $T_{n}$ is set as $\mathrm{log}(n)$. Meanwhile, if one ignores step 4, then respective method can not deal with some of the empirical networks (such as SNAP ego-networks) in Section \ref{SNAP}. 
	\item In the VH step, unlike applying K-medians in OCCAM or vertex hunting algorithm in Mixed-SCORE for hunting the $K$ centers (also known as vertices) of $\hat{R}_{*}^{*}$, we state that it is enough for our Mixed-SCORE+ to apply K-means in the VH step, and it performs satisfactory both numerically and empirically. Actually, one can also apply the VH algorithm in Mixed-SCORE or the K-medians technique in OCCAM to find the $K$ centers in Mixed-SCORE+, in this paper we use K-means.
	\item In step 6, we need to construct $\hat{V}_{*}$ and $\hat{R}^{*}_{*}$ by adding one columns with entries 1 to $\hat{V}$ and $\hat{R}^{*}$, respectively. Actually, there is a similar procedure in the MR step of Mixed-SCORE, and such procedure is related with the convex linear combination stated in \cite{mixedSCORE}.
	\item In the MR step, setting $\hat{Y}=\hat{R}^{*}_{*}\hat{V}_{*}'(\hat{V}_{*}\hat{V}_{*}')^{-1}$ in our Mixed-SCORE+ guarantees that it can deal with weak signal networks since $\hat{V}_{*}\hat{V}_{*}'$ is a $K\times K$ nonsingular matrix when $K$ is much smaller than $n$. Meanwhile, if simply setting $\hat{Y}$ as $\hat{R}^{*}\hat{V}'(\hat{V}\hat{V}')^{-1}$, then method designed based on such setting performs poor and can not successfully detect empirical networks used in this paper.
	\item In the MR step, steps 8 and 9 guarantee that $\|\hat{Y}_{i}\|$ is nonzero and all entries of $\hat{Y}_{i}$ are nonnegative (and at least one entry is strictly positive) for any $i \in \{1,2,\ldots,n\}$. This two steps make sure that $\hat{\pi}_{i}$ is well defined and nonnegative (since weights should be nonnegative for any node).
\end{itemize}
However, it is challenging to provide the respective theoretical guarantees of Mixed-SCORE+ under the degree-corrected mixed membership (DCMM) model, and we leave it for our future work.

\section{Simulations}\label{sec3}
We investigate the performance of our Mixed-SCORE+ by comparing it with Mixed-SCORE \citep{mixedSCORE}, GeoNMF \citep{GeoNMF}, SPACL \citep{SPACL} and OCCAM \citep{OCCAM} on various simulations in this section.
Note that in this paper, we only compare our Mixed-SCORE+ with methods designed for mixed membership community detection problem. It is not our intention to compare Mixed-SCORE+ with community detection methods such as those applied in \cite{SCORE+}.


For each method, we measure the performance of mixed membership community detection method by the mixed-Hamming error rate which is defined as
\begin{align*}
	\mathrm{min}_{O\in\{ K\times K\mathrm{~permutation~matrix}\}}\frac{1}{n}\|\hat{\Pi}O-\Pi\|_{1}.
\end{align*}
where $\Pi$ and $\hat{\Pi}$ are the true and estimated mixed membership matrices respectively. For simplicity, we write the mixed-Hamming error rate as $\sum_{i=1}^{n}\|\hat{\pi}_{i}-\pi_{i}\|_{1}/n$. For all the experiments in this section, we always report the mean of the mixed-Hamming error rates for every approaches, therefore for all the figures in this section, the y-axis always records the mean of $\sum_{i=1}^{n}\|\hat{\pi}_{i}-\pi_{i}\|_{1}/n$.

Unless specified, for all experiments, we set $n=500$ and $K=3$. For $0\leq n_{0}\leq 160$, let each block own $n_{0}$ number of pure nodes. For the top $3n_{0}$ nodes $\{1,2, \ldots, 3n_{0}\}$, we let these nodes be pure and let nodes $\{3n_{0}+1, 3n_{0}+2,\ldots, 500\}$ be mixed. Fixing $x\in [0, \dfrac{1}{2})$, let all the mixed nodes have four different memberships $(x, x, 1-2x), (x, 1-2x, x), (1-2x, x, x)$ and $(1/3,1/3,1/3)$, each with $\dfrac{500-3n_{0}}{4}$ number of nodes. Fixing $\rho\in(0, 1)$, the mixing matrix $P$ has diagonals 0.8 and off-diagonals $\rho$. There are two settings about $\theta$, one is $\theta(i)=0.2+0.8(i/n)^{2}$; the other is: fix $z\geq 1$, generate the degree parameters such that $1/\theta(i)\overset{iid}{\sim}U(1,z)$, where $U(1,z)$ denotes the uniform distribution on $[1, z]$. For  each parameter setting, we report the mixed-Hamming error rate $\sum_{i=1}^{n}\|\hat{\pi}_{i}-\pi_{i}\|_{1}/n$ averaged over 50 repetitions. Based on these settings we designed four experiments to illustrate the proposed method from different aspects.

\begin{figure}
	\centering
	\subfigure[Experiment 1(a)]{
		\includegraphics[width=0.45\textwidth]{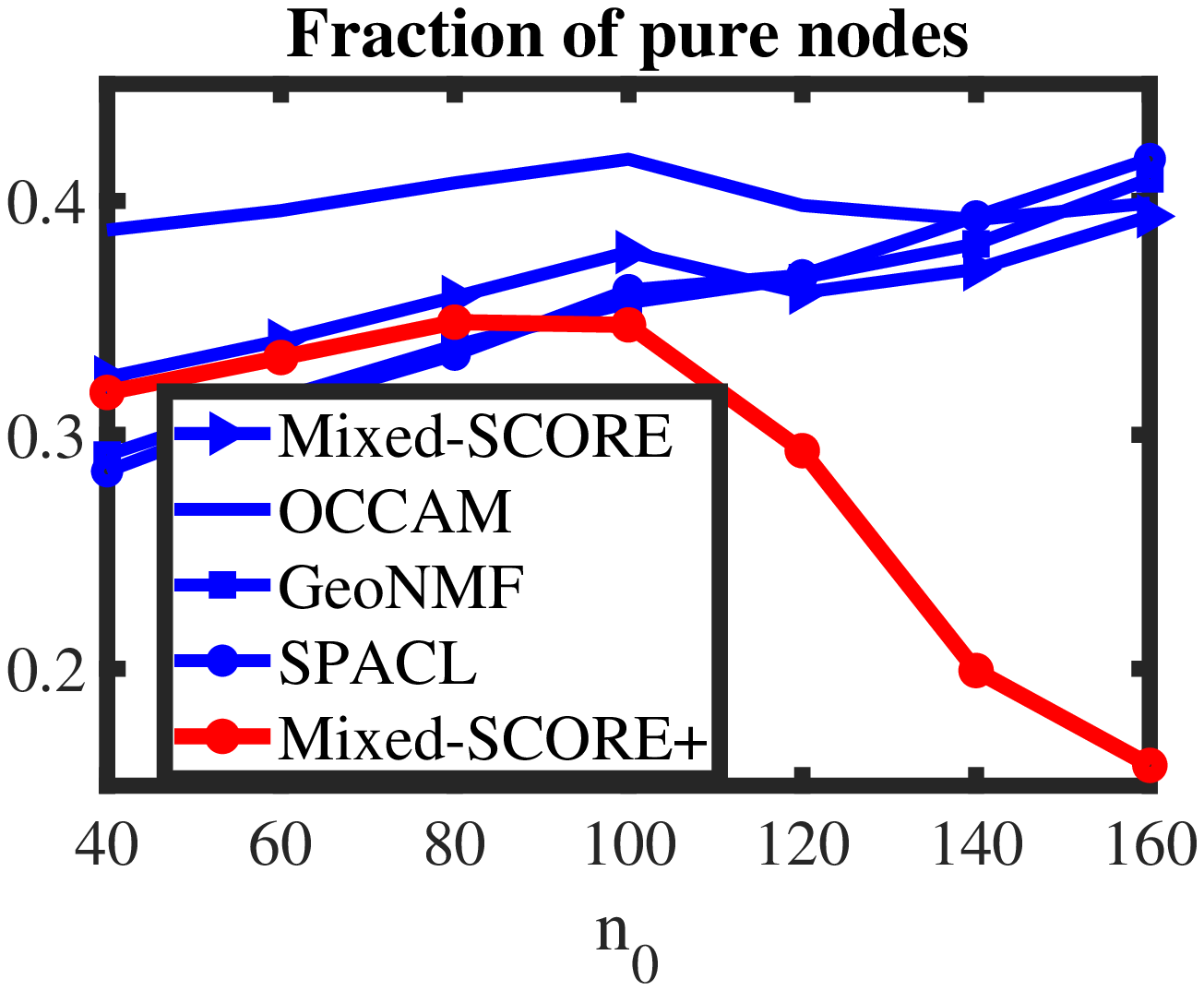}
	}
	\subfigure[Experiment 1(b)]{
		\includegraphics[width=0.45\textwidth]{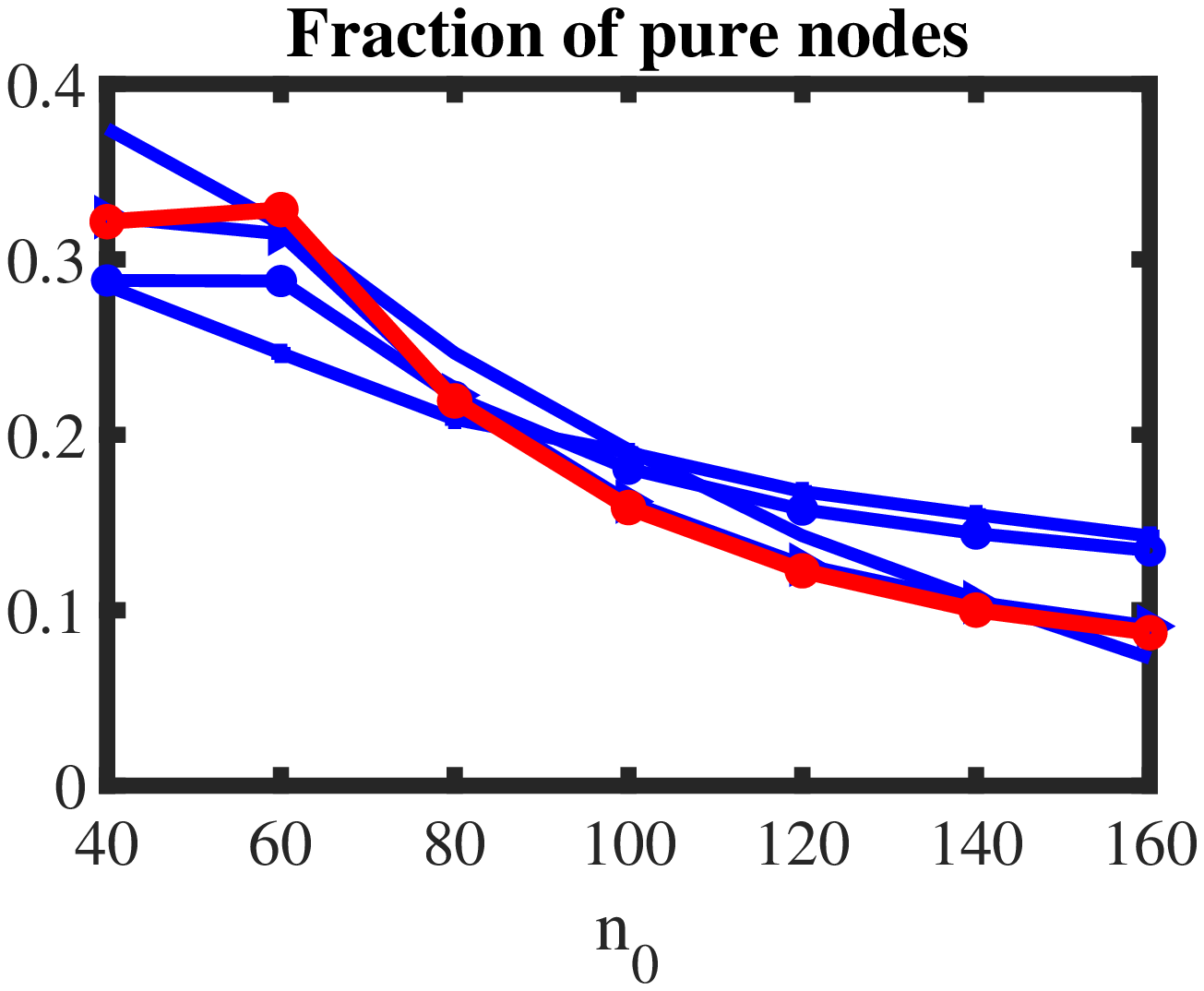}
	}
	\caption{Estimation errors of Experiments 1 (y-axis: $\sum_{i=1}^{n}n^{-1}\|\hat{\pi}_{i}-\pi_{i}\|_{1}$).}
	\label{EX1}
\end{figure}

\texttt{Experiment 1: Fraction of pure nodes.} Fix $(x,\rho)=(0.4, 0.3)$ and let $n_{0}$ range in $\{40, 60, 80, 100, 120, 140, 160\}$. A larger $n_{0}$ indicates a case with higher fraction of pure nodes. In Experiment 1(a), set $\theta(i)=0.2+0.8(i/n)^{2}$. In Experiment 1(b), set $z=4$. The numerical results are shown in panels (a) and (b) of Figure \ref{EX1}, from which we can find that all methods perform poor when the fraction of pure nodes is small. Under the setting of Experiment 1(a), our Mixed-SCORE+ significantly outperforms its competitors, and it is interesting to find that Mixed-SCORE, OCCAM, GeoNMF and SPACL always perfrom unsatisfactory under this setting even when $n_{0}$ is quite large. For Experiment 1(b), Mixed-SCORE+ performs similar as Mixed-SCORE and both two algorithms outperform OCCAM, GeoNMF and SPACL.

\begin{figure}
	\centering
	\subfigure[Experiment 2(a)]{
		\includegraphics[width=0.45\textwidth]{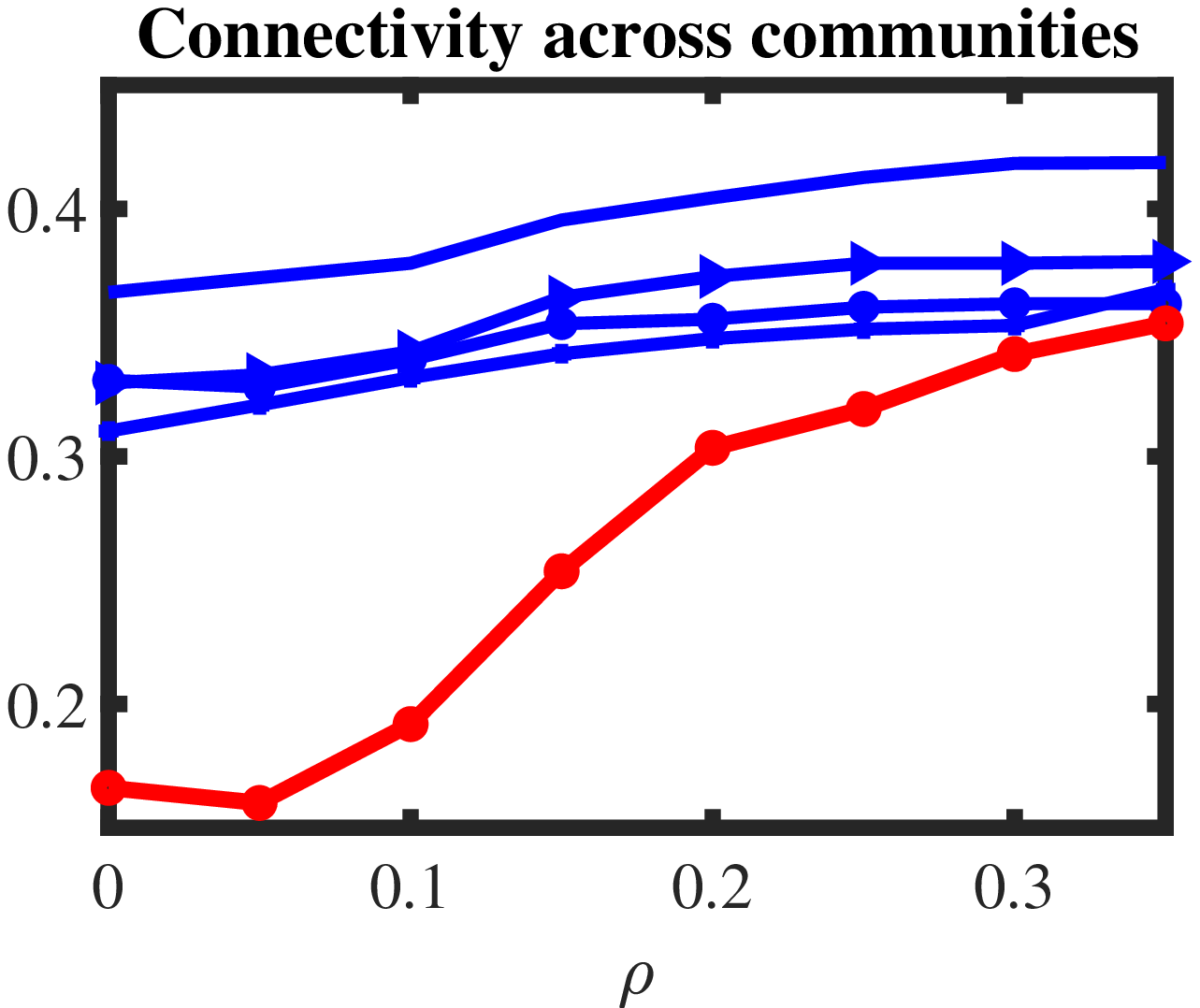}
	}
	\subfigure[Experiment 2(b)]{
		\includegraphics[width=0.45\textwidth]{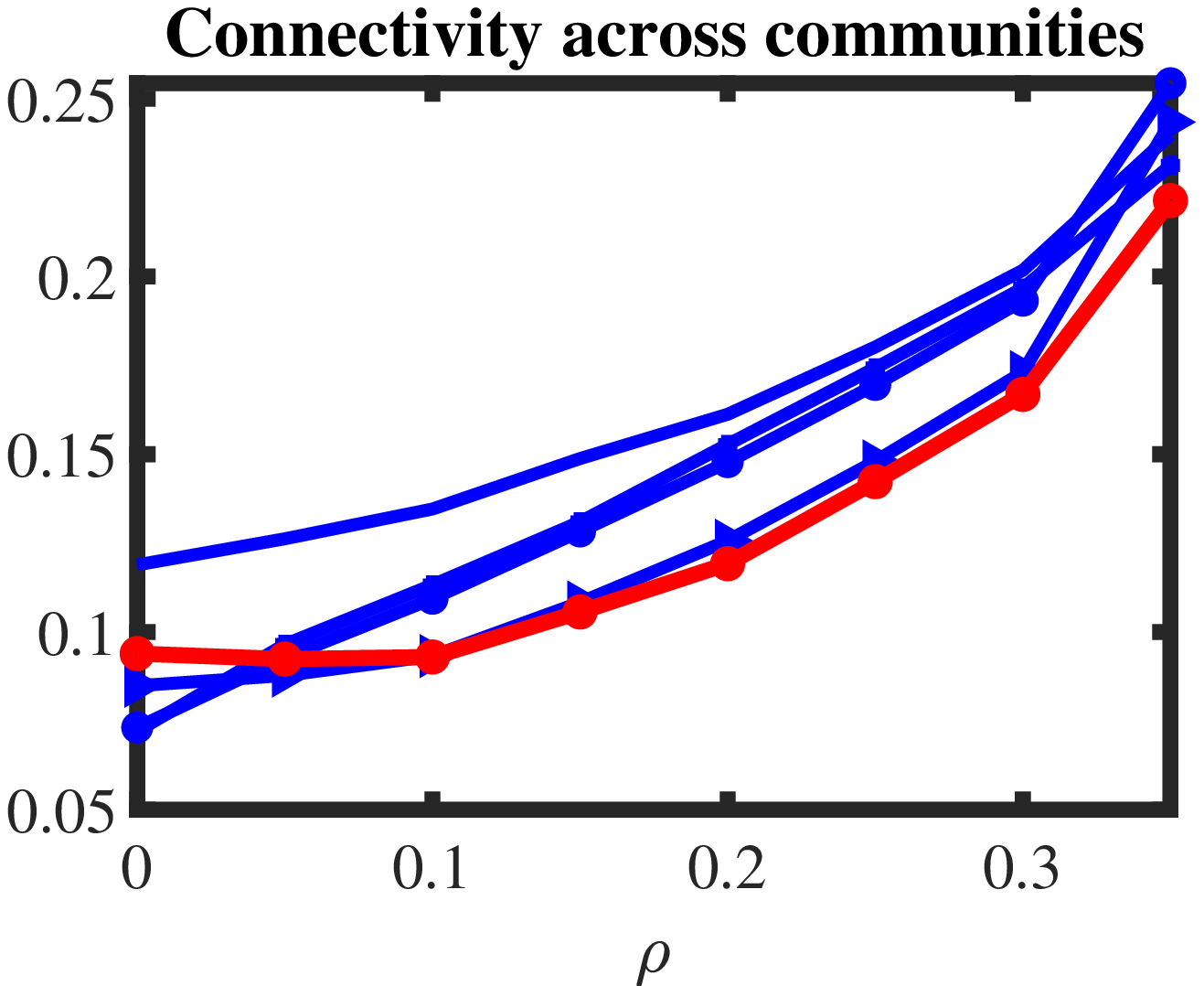}
	}
	\caption{Estimation errors of Experiments 2 (y-axis: $\sum_{i=1}^{n}n^{-1}\|\hat{\pi}_{i}-\pi_{i}\|_{1}$).}
	\label{EX2}
\end{figure}

\texttt{Experiment 2: Connectivity across communities.} Fix $(x,n_{0})=(0.4, 100)$ and let $\rho$ range in $\{0, 0.05, 0.1, \ldots, 0.35\}$. A lager $\rho$ generate more edges across different communities (hence a dense network). In Experiment 2(a), set $\theta(i)=0.2+0.8(i/n)^{2}$. In Experiment 2(b), set $z=4$. The results are displayed in  Figure \ref{EX2}. We can find that all methods perform poorer as $\rho$ increases, this phenomenon occurs due to the fact that more edges across different communities lead to a case that different communities tend to be into a giant community and hence a case that is more challenging to detect for any algorithms. Under the setting of Experiment 2(a), our Mixed-SCORE+ outperforms its competitors obviously, and the 4 competitors always perform poor even for a small $\rho$. Meanwhile, in Experiment 2(b), Mixed-SCORE+ performs slightly better than Mixed-SCORE while both two approaches outperform OCCAM, GeoNMF and SPACL.

\begin{figure}[!h]
	\centering
	\subfigure[Experiment 3(a)]{
		\includegraphics[width=0.45\textwidth]{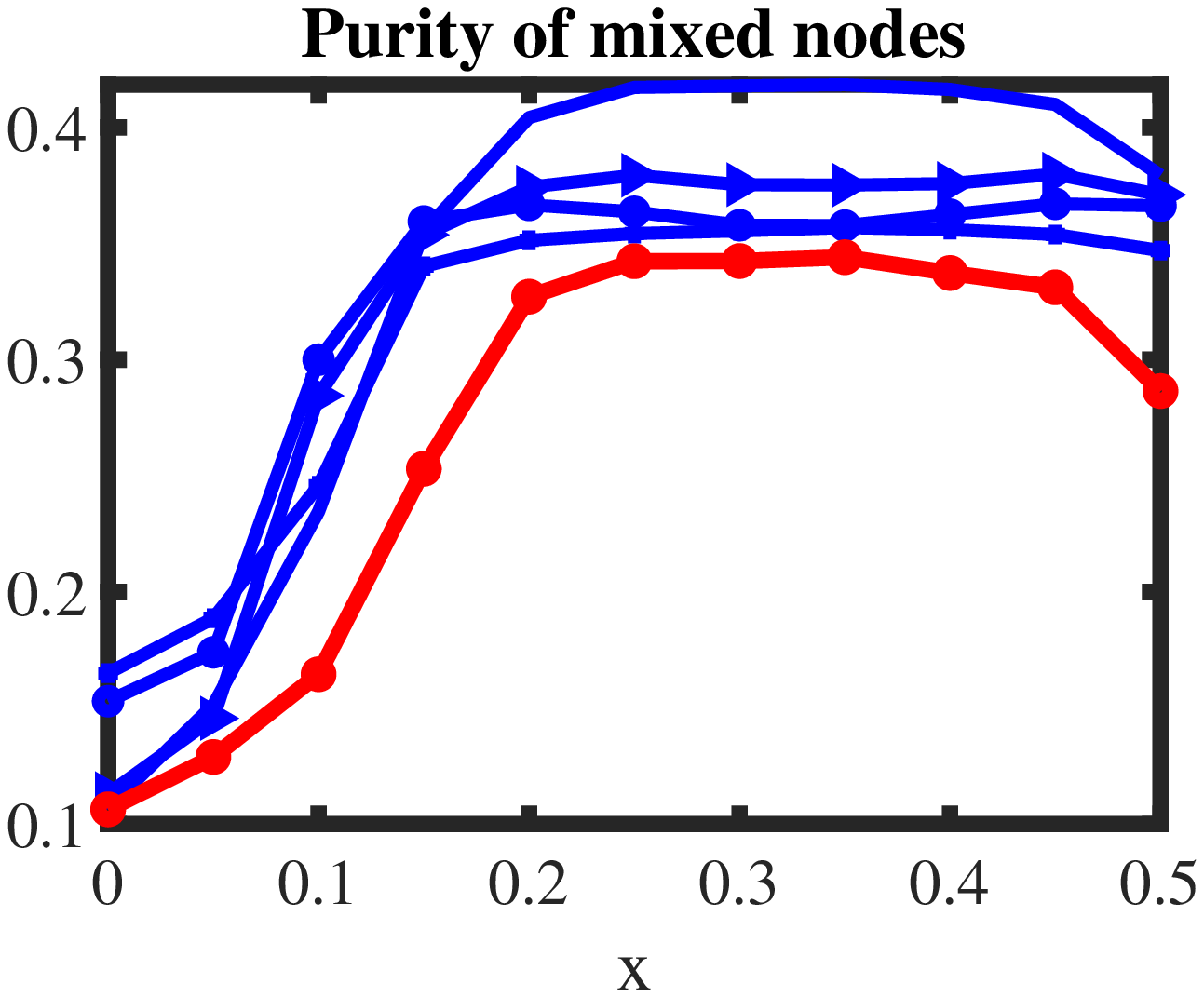}
	}
	\subfigure[Experiment 3(b)]{
		\includegraphics[width=0.45\textwidth]{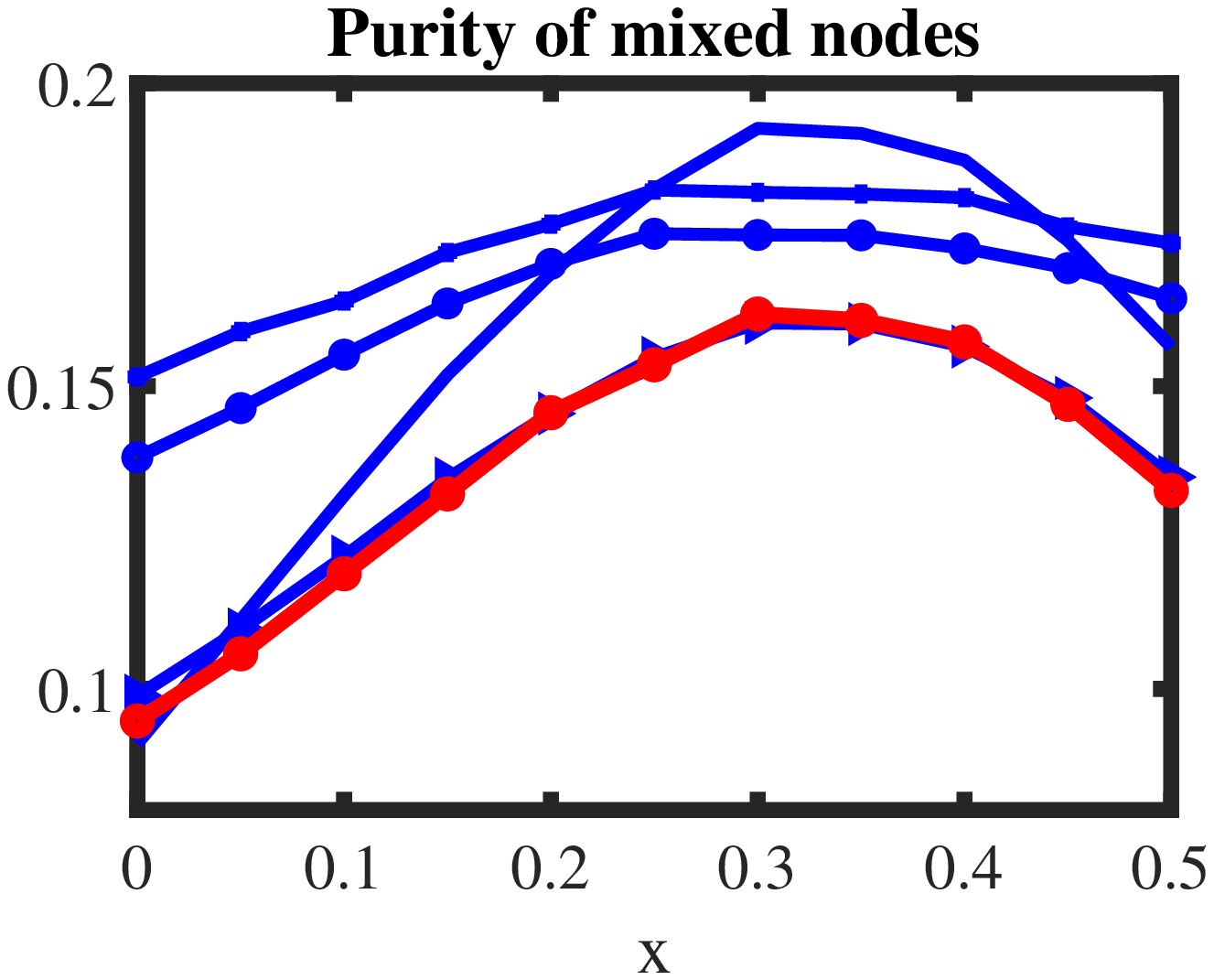}
	}
	\caption{Estimation errors of Experiments 3 (y-axis: $\sum_{i=1}^{n}n^{-1}\|\hat{\pi}_{i}-\pi_{i}\|_{1}$).}
	\label{EX3}
\end{figure}

\texttt{Experiment 3: Purity of mixed nodes.} Fix $(n_{0},\rho)=(100, 0.3)$, and let $x$ range in $\{0, 0.05, \ldots, 0.5\}$. As $x$ increases to 1/3, these mixed nodes become less pure and they become more pure as $x$ increases further. In Experiment 3(a), set $\theta(i)=0.2+0.8(i/n)^{2}$. In Experiment 3(b), set $z=4$. Figure \ref{EX3} records the numerical results of this experiment. It is obvious to find that Mixed-SCORE+ outperforms the other four methods in Experiment 3(a), and it performs similar as Mixed-SCORE+ while both two perform better than OCCAM, GeoNMF and SPACL.

\begin{figure}
	\centering
	\subfigure[Experiment 4(a)]{
		\includegraphics[width=0.45\textwidth]{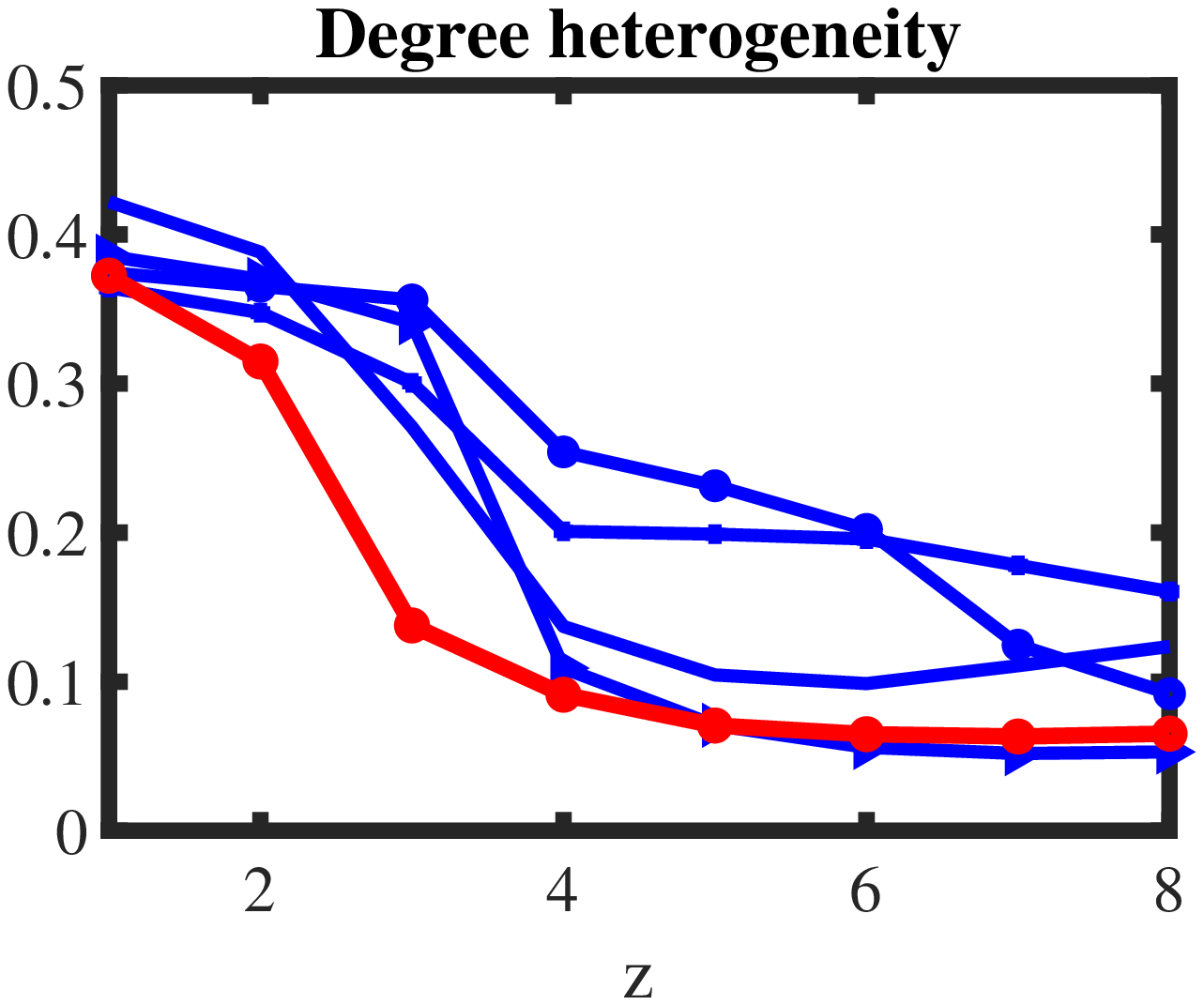}
	}
	\subfigure[Experiment 4(b)]{
		\includegraphics[width=0.45\textwidth]{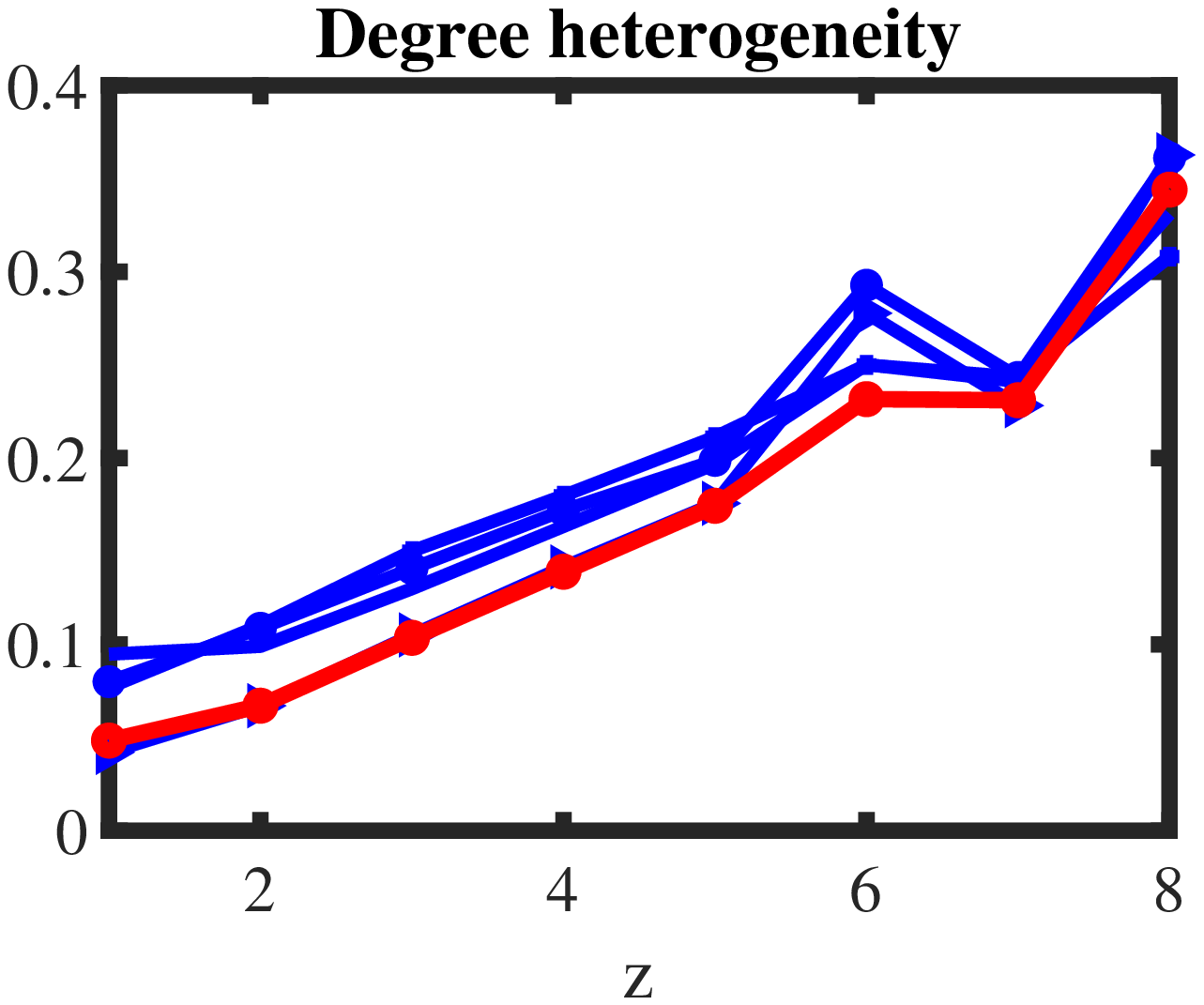}
	}
	\caption{Estimation errors of Experiments 4 (y-axis: $\sum_{i=1}^{n}n^{-1}\|\hat{\pi}_{i}-\pi_{i}\|_{1}$).}
	\label{EX4}
\end{figure}

\texttt{Experiment 4: Degree heterogeneity.} Fix $(n_{0},\rho, x)=(100, 0.3, 0.4)$. and let $z$ range in $\{1,2,\ldots,8\}$. In Experiment 4(a), set $\theta(i)=z/10+0.8*(i/n)^{2}$. In Experiment 4(b), set $1/\theta(i)\overset{iid}{\sim}U(1,z), i=1,2,\ldots,n$. From the results in Figure \ref{EX4} we can conclude that this experiment shares similar conclusions with the above experiments.

\section{Application to empirical datasets}\label{sec4}
In this section, we apply two kinds of empirical datasets to investigate the performance of our Mixed-SCORE+. For the community detection problem, we  use the eight real-world networks with known label information; for the mixed membership community detection problem, we use the SNAP ego-networks with known membership information where the SNAP ego-networks are applied in \cite{OCCAM} and \cite{MixedRSC}.

\subsection{Eight empirical networks with known label information for community detection}
The details of the eight real-world networks can be found in Appendix \ref{dereal8}. To measure the performances of these methods on the eight networks, we first introduce the Hamming error rate.

When all nodes are pure,  the community information can be expressed by an $n\times 1$ nodes labels vector $\ell$ where $\ell_i$ takes values in set $\{1, \cdots, K\}$ and  denotes  the node $i$ belongs to the $\ell_i$-th community. Let $\hat{\ell}$ be an estimation of $\ell_i$. 
For community detection, since each node belongs to exactly one community, $\hat{\ell}_i$ and $\ell_i$ take one value from $\{1,2, \ldots, K\}$ for $1\leq i\leq n$. $\hat{\ell}$ for Mixed-SCORE+ can be computed as below
\begin{align*}
	\hat{\ell}_i=\mathrm{arg~max}_{1\leq k\leq K}\hat{\Pi}_{ik}.
\end{align*}
Then the clustering error rate is measured by the Hamming error rate \citep{SCORE} which is defined as
\begin{align*}
	\mathrm{min}_{\{o:\mathrm{~permutation~over~}\{1,2,\ldots, K\}\}}\frac{1}{n}\sum_{i=1}^{n}1\{o(\hat{\ell}_{i})\neq \ell_{i}\},
\end{align*}
where $\ell_{i}$ and $\hat{\ell}_{i}$ are the true and estimated labels of node $i$.

The error rates of the eight empirical networks are summarized in Table \ref{real8errors}, where we use default parameters for Mixed-SCORE+. The results show that Mixed-SCORE+ outperforms its competitors on the three large networks: Polblogs, Simmons and Caltech, with error rates 54/1222, 125/1137, and 94/590 respectively. As discussed in \cite{SCORE+}, Simmons and Caltech are two weak signal \footnote{Readers interested in the details of weak signal networks and strong signal networks please refer to \cite{SCORE+}.} networks whose $(K+1)$-th leading eigenvalue is close to the $K$-th leading eigenvalue of the adjacency matrix $A$ or its variants, suggesting that the leading $(K+1)$ eigenvector may contain information about nodes labels.
While, for the five small strong signal networks, we see that all methods enjoy similar performances.
\begin{table}[h!]
	\footnotesize
	\centering
	\caption{Error rates on the eight empirical data sets.}
	\label{real8errors}
	\resizebox{\columnwidth}{!}{
		\begin{tabular}{cccccccccc}
			\toprule
			\textbf{ Methods} &Karate&Dolphins&Football&Polbooks&UKfaculty&Polblogs&Simmons&Caltech\\
			\midrule
			Mixed-SCORE&\textbf{0/34}&2/62&\textbf{4/110}&3/92&6/79&60/1222&261/1137&174/590\\
			OCCAM&\textbf{0/34}&\textbf{1/62}&\textbf{4/110}&3/92&5/79&60/1222&268/1137&192/590\\
			GeoNMF&\textbf{0/34}&\textbf{1/62}&5/110&3/92&4/79&64/1222&383/1137&229/590\\
			SPACL& \textbf{0/34}&\textbf{1/62}&5/110&3/92&4/79& 61/1222&413/1137 & 185/590\\
			\hline
			Mixed-SCORE+&1/34&\textbf{1/62}&6/110&\textbf{2/92}&2/79&\textbf{54/1222}&\textbf{125/1137}&\textbf{94/590}\\
			\bottomrule
	\end{tabular}}
\end{table}
\subsection{SNAP ego-networks with known mixed membership information for mixed membership community detection}\label{SNAP}
SNAP ego-networks contains substantial ego-networks from three platforms Facebook, GooglePlus, and Twitter. There are 7 communities with total 1656 nodes in Facebook, 58 communities with total 25127 nodes in GooglePlus, and 255 communities with total 15463 nodes in Twitter.  For more details please refer to \cite{OCCAM} and \cite{MixedRSC}. Here we use the newest version of SNAP ego-networks (those used in \cite{MixedRSC}) to investigate the performances of Mixed-SCORE+ and its competitors.

Since the ground truth communities of mixed membership (i.e., $\Pi$) of SNAP ego-networks are known in advance, we can use the mixed-Hamming error rate to measure these methods' performances directly. To compare the performances of these methods, similar as that in \cite{OCCAM}, we report the average performances over each of the social platforms and the corresponding standard deviation in Table \ref{ErrorSNAP}. Meanwhile, recall that in the VH step of Mixed-SCORE+, we argue that we apply K-means method for vertices hunting instead of K-medians. Here, we use Mixed-SCORE+(Kmedians) to denote the Mixed-SCORE+ method designed based on K-medians clustering technique. We also report the numerical results of Mixed-SCORE+(Kmedians) on the SNAP ego-networks in Table \ref{ErrorSNAP}, which tells us that Mixed-SCORE+ shares similar \footnote{Actually, Mixed-SCORE+ also shares almost the same error rates as that of Mixed-SCORE+(Kmedians) on the eight real-world networks in Tabel \ref{real8errors}.} performances as that of Mixed-SCORE+(Kmedians). Since K-means is faster than K-medians, the default vertices hunting technique for Mixed-SCORE+ is K-means in this paper. From Table \ref{ErrorSNAP}, we can find that, for Facebook networks,  SPACL has smallest error rate, while GeoNMF and Mixed-SCORE+ have similar results. OCCAM performs poorest on Facebook networks. For GooglePlus and Twitter networks, our proposed methods Mixed-SCORE+ and Mixed-SCORE+(Kmedians) perform best and share similar error rates. At the same time, we see that Mixed-SCORE performs poorest on GooglePlus and Twitter, suggesting that our Mixed-SCORE+ provides a significant improvement of Mixed-SCORE.
\begin{table}[h!]
	\centering
	\caption{Mean (SD) of mixed-Hamming error rates for ego-networks.}
	\label{ErrorSNAP}
	\resizebox{\columnwidth}{!}{
		\begin{tabular}{cccccccccc}
			\toprule
			&Facebook&GooglePlus&Twitter\\
			\midrule
			Mixed-SCORE&0.2496(0.1322)&0.3766(0.1053)&0.3088(0.1296)\\
			OCCAM&0.2610(0.1367)&0.3564(0.1210)&0.2864(0.1406)\\
			GeoNMF&0.2537(0.1266)&0.3520(0.1078)&0.2858(0.1292)\\
			SPACL& \textbf{0.2371}(0.1233)&0.3616(0.1077)& 0.3068(0.1268)\\
			\hline
			Mixed-SCORE+&0.2536(0.1289)&0.3341(0.1157)&\textbf{0.2659}(0.1411)\\
			Mixed-SCORE+(Kmedians)&0.2561(0.1292)&\textbf{0.3332}(0.1168)&0.2665(0.1422)\\
			\bottomrule
	\end{tabular}}
\end{table}
\section{Discussion}\label{sec5}
In this paper,  Mixed-SCORE+ focus on detecting network memberships for the problem of mixed membership community detection, and it can also  detect two weak signal networks Simmons and Caltech. Such advantage of Mixed-SCORE+ mainly comes from the fact we apply the information of the leading $(K+1)$ eigenvector and eigenvalue of the regularized Laplacian matrix when dealing with weak signal networks. Although Mixed-SCORE+ is an extension of Mixed-SCORE,  Mixed-SCORE can not utilize such information (for the details, please refer to those remarks after our Mixed-SCORE+ algorithm). Numerical studies of substantial simulations and empirical datasets show that Mixed-SCORE+ enjoys satisfactory performances and it performs better than most of the benchmark methods both numerically and empirically.

There remain several problems unsolved: (a) \cite{mixedSCORE} provided full theoretical analysis for Mixed-SCORE while there is no such studies for Mixed-SCORE+ in this paper due to the fact that it is challenge and difficult to study the theoretical guarantee of Mixed-SCORE+. Hence, it is meaningful to build theoretical frameworks for Mixed-SCORE+. (b)  Whether there exist optimal parameters $\tau$ and $T_{n}$ both theoretically and numerically is an interesting topic for further study. (c) In \cite{ali2018improved}, the authors studied the existence of an optimal value $\alpha_{opt}$ of the parameter $\alpha$ for community detection methods based on $D^{-\alpha}AD^{-\alpha}$ for community detection problem. Recall that our Mixed-SCORE+ is designed based on $D_{\tau}^{-1/2}AD_{\tau}^{-1/2}$, we argue that whether there exist optimal $\alpha_{0}$ and $\beta_{0}$ such that mixed membership community detection method (say Mixed-SCORE+) designed based on $D^{\alpha_{0}}_{\tau}A^{\beta_{0}}D^{\alpha_{0}}_{\tau}$ outperforms methods designed based on $D^{\alpha}_{\tau}A^{\beta}D^{\alpha}_{\tau}$ for any choices of $\alpha$ and $\beta$. For reasons of space, we leave studies of these problems to the future.

\appendix
\section{ Description of eight real-word data}\label{dereal8}
\begin{itemize}
	\item \textbf{Karate}: this network consists of 34 nodes where each
	node denotes a member in the karate club \citep{karate}. As there is a conflict in the club, the network divides into two communities: Mr. Hi’s group and John’s group. \cite{karate} records all labels for each member and we use them as the true labels.
	\item \textbf{Dolphins}: this network consists of frequent associations between 62 dolphins in a community living off Doubtful Sound. In Dolphins network, node denotes a dolphin, and edge stands for companionship \citep{dolphins0, dolphins1, dolphins2}. The network splits naturally into two large groups females and males \citep{dolphins1, dolphinnewman}, which are seen as the ground truth in our analysis.
	\item \textbf{Football}: this network is for American
	football games between Division I-A college teams during the regular football season of Fall \citep{football}. Nodes in Football denote teams and edges represent regular-season games between any two teams \citep{football}. The original network contains 115 nodes in total, since 5 of them are called ``Independent'' and the remaining 110 nodes are manually divided into 11 conferences for administration purpose, for community detection, we remove the 5 independent teams in this paper.
	\item \textbf{Polbooks}: this network is about US politics
	published around the 2004 presidential election and sold by the online bookseller Amazon.com. In Polbooks, nodes represent books, edges represent frequent co-purchasing of books by the same buyers. Full information about edges and labels can be downloaded from \url{http://www-personal.umich.edu/~mejn/netdata/}. The original network contains 105 nodes labeled as either ``Conservative”, ``Liberal”, or ``Neutral”. Nodes labeled ``Neutral'' are removed for community detection in this paper.
	\item \textbf{UKfaculty}: this network reflects the friendship among academic staffs of a given Faculty in a UK university consisting of three separate schools \citep{UKfaculty}. The original network contains 81 nodes, in which the smallest group only has 2 nodes. The smallest group is removed for community detection in this paper.
	\item \textbf{Polblogs}: this network consists of political blogs during the
	2004 US presidential election \citep{Polblogs1}. Each blog belongs to one of the two parties liberal or
	conservative. As suggested by \cite{DCSBM}, we only consider the largest connected component with 1222  nodes and ignore the edge direction for community detection.
	\item \textbf{Simmons}: this network contains one largest connected
	component with 1137 nodes. It is observed in \cite{traud2011comparing, traud2012social} that the community structure of the Simmons College network exhibits a strong correlation with the graduation year-students since students in the same year are more likely to be friends.
	\item \textbf{Caltech}: this network has one largest connected component with 590 nodes. The community structure is highly correlated with which of the 8 dorms a user is from, as observed in \cite{traud2011comparing, traud2012social}.
\end{itemize}
\bibliographystyle{Chicago}
\bibliography{reference}
\end{document}